\documentclass[twocolumn,amsmath,amssymb,longbibliography,aps,prl,reprint,
superscriptaddress]{revtex4-1}
\usepackage{graphicx}
\usepackage{amsmath}
\usepackage{bm}
\usepackage{comment}
\usepackage{multirow}
\usepackage{color}
\usepackage{amsfonts}
\usepackage{booktabs}
\usepackage{ulem}
\usepackage{siunitx}
\newcommand{\red}{\color{red}}
\newcommand{\blue}{\color{blue}}

\begin{document}

\title{Parity and time-reversal invariant Ising spin ordering}
\title{Intrinsic longitudinal spin-conductivity from ferroaxial spin-symmetry breaking}

\author{Yue Yu}
\affiliation{Department of Physics, University of Wisconsin--Milwaukee, Milwaukee, Wisconsin 53201, USA} 
\affiliation{School of Physics and Astronomy, University of Minnesota, Minneapolis, Minnesota 55455, USA} 

\author{Jin Matsuda}
\affiliation{Department of Physics, The University of Tokyo, Hongo, Bunkyo-ku, Tokyo 113-0033, Japan}

\author{Hikaru Watanabe}
\affiliation{Graduate School of Engineering, Hokkaido University, Sapporo 060-8628, Japan}

\author{Ryotaro Arita}
\affiliation{Department of Physics, The University of Tokyo, Hongo, Bunkyo-ku, Tokyo 113-0033, Japan}
\affiliation{RIKEN Center for Emergent Matter Science, 2-1 Hirosawa, Wako, Saitama 351-0198, Japan}

\author{Daniel F. Agterberg}
\affiliation{Department of Physics, University of Wisconsin--Milwaukee, Milwaukee, Wisconsin 53201, USA} 
\begin{abstract}
The interplay of antiferromagnetic order, momentum-dependent Bloch spin-splitting, time-reversal (T), and parity (P) symmetries in non-relativistic systems has emerged as a central theme for spintronics. Two well-known examples are P-preserving and T-violating altermagnets and P-violating and T-preserving odd-parity magnets. Here we examine a class of coplanar AFMs with a translation invariant vector spin chiral (VSC) order that preserves both P and T symmetries but breaks the non-relativistic spin-rotational invariance. Naively, such a VSC order is not expected to exhibit unusual phenomena. Here we show that the spin-rotational symmetry breaking generated by this VSC order can take on multiple forms, including ferroaxial symmetry breaking. We further show that these states allow for non-relativistic and non-dissipative purely longitudinal (or purely transverse) spin-conductivities.  These states also allow the generation of non-relativistic altermagnetic spin-splittings through circularly polarized light. We identify 16 candidate materials in the Magndata database for which our theory is applicable and provide effective microscopic models and DFT-based results that highlight the large emergent spin conductivities.

\end{abstract}
\maketitle

{\it Introduction: }
Altermagnetism has recently emerged as a central topic in quantum magnetism\cite{hayami2019,yuan2020,mazin2021,vsmejkal2022,vsmejkal2022emerging}. Like ferromagnetism, it exhibits collinear spin ordering, preserves lattice translation, breaks time-reversal $T$ symmetry, and preserves inversion $P$ symmetry. The $(P,T)=(+,-)$ property leads to nonrelativistic even-parity  spin splittings in momentum space\cite{krempasky2024}, that enables spin-transport\cite{gonzalez2021,han2024} and spin caloritronics\cite{cui2023}. 
When spin-orbit coupling (SOC) is present, altermagnets can also generate anomalous Hall transport\cite{vsmejkal2020,Roig:2025}. 

 Driven by altermagnetism, interest in other unconventional magnets has also grown\cite{jiang2024,xiao2024,chen2024,watanabe2024Symmetry}, notably odd-parity (or $p$-wave) magnets\cite{hellenes2023,matsuda2024,brekke2024,yu2025,luo2025}. These antiferromagnets (AFMs) break lattice translational symmetry with coplanar spin orderings. This coplanar spin order induces a vector spin chirality (VSC), $\vec{S}_1\times \vec{S}_2$,   that is translation invariant and is $(P,T)=(-,+)$, allowing odd-parity Bloch spin splittings where the spin in momentum space is collinear, or Ising-like \cite{hellenes2023,brekke2024,yu2025}. Recently, non-Ising-like odd-parity spin-splittings have been examined in non-collinear AFMs \cite{luo2025}. 
As nonrelativistic analogues of Ising and Rashba SOC,  they can play a significant role in spintronics\cite{manchon2015,gonzalez2024,hu2024,chakraborty2025}. 
 
PT-symmetric magnets, with $(P,T)=(-,-)$, have also received renewed scrutiny \cite{watanabe2020,watanabe2024,hayami2022}. The $PT$ symmetry guaranties doubly-degenerate bands, hiding the spin-charge locking\cite{zhang2014}. Bloch spin splittings can be unlocked by spin-independent perturbations that break $P$ or $T$. Ferroelectricity, electric fields, or surface $P$-breaking can generate even-parity ferromagnetic\cite{mazin2023} or altermagnetic\cite{mazin2023,wang2024,gu2025,sun2025,mavani2025,Smejkal2024altermagneticME,Duan2025AFAM} spin splittings. Circularly polarized light can generate odd-parity Ising spin splittings\cite{Zhou2025,zhu2025,li2025,huang2025}.

In this work, we re-examine the final class of spin-orders that are even under both inversion and time-reversal symmetries: $(P,T)=(+,+)$. 
Generally, $(P,T)=(+,+)$ symmetric order, like $(P,T)=(-,-)$ order, forbids spin-splittings in momentum space. This suggests that such an ordering will be uninteresting from a spintronics perspective. Here we argue that this is not the case. In particular, we examine lattice doubling coplanar AFMs that induce an {\it even-parity} spatially-uniform vector spin chirality (VSC). Since its importance was pointed out by Villain in 1977 \cite{villain1977}, VSC has been shown to play an important role in quantum materials\cite{kawamura1984,katsura2005,yu2025}. From the spintronics perspective, odd-parity VSC is central to the generation of both electric polarization \cite{katsura2005} and odd-parity magnetism \cite{hellenes2023,yu2025,brekke2024}.  

The role of even parity VSC in spintronics has received less attention. Here we classify the even-parity AFM and argue that a spatially uniform even-parity VSC drives a spin-rotational symmetry breaking that allows dissipationless spin conductivities. Such spin-conductivities have traditionally been understood as a consequence of SOC \cite{dyakonov1971,sinova2015,kato2004}. Here, they emerge in a non-relativistically and hence can be parametrically larger. In addition, we find that the spatial symmetry of the even-parity VSC can lead to unusual forms of spin conductivity. In particular, we find that two types of VSC appear most frequently in  our symmetry-based classification. The first type gives rise to purely transverse, or spin-Hall conductivities. Here these spin-Hall conductivities are generated non-relativistically, but they are also commonly generated by SOC. The second is a ferroaxial VSC, which gives rise to purely longitudinal spin-conductivities. Notably, purely longitudinal spin-conductivities in time-reversal preserving materials have not been reported elsewhere. Such dissipationless longitudinal spin conductivities can be generated by SOC in low-symmetry crystal structures\cite{Wimmer:2015,Roy2022,hayami2022spin}, but symmetry requires they must coexist with spin-Hall conductivities which are typically larger in magnitude.

It is worth emphasizing that, in contrast to other emergent spin current responses, such as in Mn$_3$Sn \cite{Zhang:2018,Zhang:2015} and altermagnetic materials~\cite{Naka2019-th,Ahn2019-cd,Ma2021-ji,Gonzalez-Hernandez2021-pb}, in which time-reversal symmetry is broken, here only time-reversal even spin responses appear, which are independent of impurity scattering \cite{Freimuth2014,Jungwirth2017}, allowing control for spintronics applications.

In the following, we present our group-theory-based microscopic
template for $(P,T)=(+,+)$ VSC induced by 166 period-doubling
AFMs and provide explicit microscopic models for these. 
We apply both phenomenological and microscopic analysis to demonstrate the non-zero transverse and longitudinal  non-relativistic spin conductivities from the $(P,T)=(+,+)$ VSC. Using first-principles calculations, we show that the non-relativistic spin Hall conductivity in U$_2$Ni$_2$In is comparable to that of Pt, providing a realistic demonstration of the utility of even parity VSC. Finally we show that $(P,T)=(+,+)$ VSC gives rise to controllable spin-splittings through the application circularly polarized light.

\begin{figure}[h]
\centering
\includegraphics[width=6.5cm]{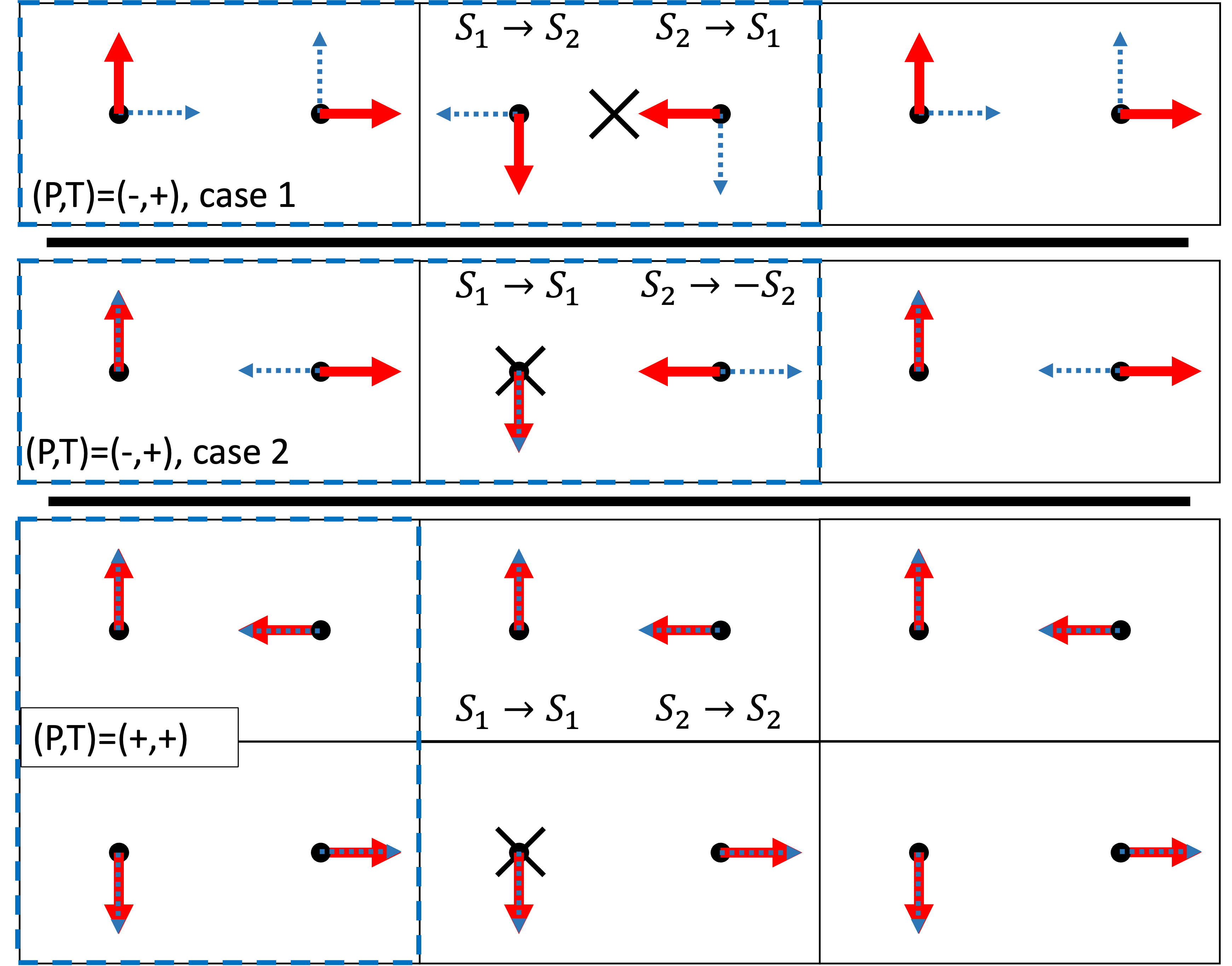}
\caption{Spin configurations of coplanar AFM states, denoted by red solid arrows. Inversion partners are illustrated by
blue dashed arrows. The inversion center is at the X mark. Black solid (blue dashed) boxes denote the nonmagnetic (magnetic) unit cell. For $(P,T)=(-,+)$ odd-parity magnets, inversion symmetry acts non-trivially on the AFM, while for $(P,T)=(+,+)$ state inversion is trivial.  }
\label{F:2}
\end{figure}

{\it Symmetry Arguments: }
Prior to presenting our symmetry-based arguments, we give a simplified example that highlights the essential ingredients. Figure \ref{F:2} illustrates coplanar period-doubling AFM states that emerge from a non-magnetic state with inversion symmetry. The nonmagnetic unit cell contains two atoms with magnetic moments ${\bf S}_{1,2}$.  Consideration of inversion symmetry naturally leads to the three cases in Fig.\ref{F:2}. In the first case, inversion interchanges ${{\bf S}_1}$ and ${{\bf S}_2}$. In the second case, inversion leaves ${{\bf S}_1}$ unchanged but flips ${{\bf S}_2}$. Both results in an inversion-odd ${{\bf S}_1} \times {{\bf S}_2}$, describing p-wave, or more generally, odd-parity magnets\cite{yu2025} with $(P,T)=(-,+)$. In contrast, for the third case, the inversion symmetry acts trivially on ${\bf S}_{1,2}$, and the resulting ${{\bf S}_1} \times {{\bf S}_2}$ is even under inversion, giving an even-parity VSC. This VSC is time-reversal invariant, as time-reversal symmetry flips both spins, resulting in a $(P,T) = (+,+)$ magnet.

Our approach is to apply a Landau theory to describe the symmetry requirements to generate a phase transition into the $(P,T) = (+,+)$ magnet. Central to the above example is the existence of two independent AFM spin degrees of freedom 
${\bf S}_{1,2}$, which are required for a non-vanishing VSC. In the non-relativistic limit, each spatial degree of freedom is represented by an independent spin-vector. Consequently, a non-relativistic VSC requires an order parameter with a space group irreducible representation (IR) with momentum $\bf{Q}$ ($\Gamma_{\bf Q}$) that is (at least) two-fold degenerate. Here we consider only two-fold degenerate IR's (higher dimensional IR's will typically generate non-coplanar magnetic states, and our emphasis here is coplanar magnetic states). Labeling the basis of spatial degrees of freedom in the IR by $\Gamma_{{\bf Q},1},\Gamma_{{\bf Q},2}$, our AFM  order parameter is then ${\bf S}_1$ and ${\bf S}_2$. In previous discussion  ${\bf S}_i$ referred to spins localized at lattice sites, but here it is more generally any magnetic order consistent with the IR.  The VSC in this case is given by ${\bf S}_{1}\times{\bf S}_2$,the VSC belongs to an IR with momentum $2{\bf Q}$, if $2{\bf Q}={\bf G}$, where ${\bf G}$ is a reciprocal lattice vector, then the VSC is spatially uniform and will allow the non-vanishing spin conductivities. Furthermore, ${\bf S}_{1}\times{\bf S}_{2}$ must then transform spatially according to a point group representation, that is, it belongs to the IR $\Gamma_\text{VSC}\otimes \Gamma^S_{A}$, where $\Gamma_\text{VSC}$ is the real-space point group IR and $\Gamma^S_A$ is the spin-axial representation.  To address which spin-conductivities are non-vanishing, we require the identification of $\Gamma_\text{VSC}$. Formally, this belongs to the IR $\Gamma_{\bf Q}\otimes_A\Gamma_{\bf Q}$, where $\otimes_A$ is the antisymmetric direct product (this encodes the antisymmetry of the cross product ${\bf S}_{1}\times{\bf S}_{2}$). 

The above considerations motivate our symmetry-based search for even parity VSC. We find all two-dimensional space group representations for TRIM points ${\bf Q}$ (where $2\bf{Q}=\bf{G}$) and require both components of $\Gamma_{\bf Q}$ to be inversion symmetric (note inversion symmetry belongs to the little co-group of ${\bf Q}$ since it is a TRIM point). For simplicity, we consider only single-Q states. It is possible to extend our analysis to include multiple-Q states if needed. This symmetry approach covers 166 candidate representations. These are listed in Table~\ref{T:1} in the Supporting Information and the symmetry of the VSC is also provided.
 
For these 2D IRs, the single continuous phase transition can be described by Landau theory \cite{elcoro2017}. The free energy for the order parameter $({\bf S}_1, {\bf S}_2)$, 
derived from all 166 representations, takes a universal form:
\begin{equation}
\begin{split}
f&=a(T)\sum_i {\bf S}_i\cdot{\bf S}_i+\beta_1(\sum_i {\bf S}_i\cdot{\bf S}_i)^2\\
&+\beta_2(2{\bf S}_1\cdot{\bf S}_2)^2+\beta_3({\bf S}_1\cdot{\bf S}_1-{\bf S}_2\cdot{\bf S}_2)^2\\&+\beta_4(2{\bf S}_1\cdot{\bf S}_2)({\bf S}_1\cdot{\bf S}_1-{\bf S}_2\cdot{\bf S}_2).
\end{split}\label{E:Landau}
\end{equation}
The quartic terms differ between IRs. In particular, $\beta_{1,2,3}$ always appear. The $\beta_4$ term appears only if the two inner products ${\bf S}_1\cdot{\bf S}_2$ and ${\bf S}_1\cdot{\bf S}_1-{\bf S}_2\cdot{\bf S}_2$ carry the same symmetry. $\beta_{1,2,3}$ are typically different; however, there exist examples when $2{\bf S}_1\cdot{\bf S}_2$ and ${\bf S}_1\cdot{\bf S}_1-{\bf S}_2\cdot{\bf S}_2$ belong to the same 2D IR. In these cases $\beta_2=\beta_3$. The ground state solution depends upon the specific values of $\beta_i$ and is discussed in more detail in the Supporting Information. Generally, by minimizing the free energy, we find two collinear AFM states and a coplanar state with nonvanishing VSC order ${{\bf S}_1} \times {{\bf S}_2}$. Having established its existence, in the following, we turn our attention to the coplanar state with non-vanishing VSC order. In Ref.\cite{Watanabe2026}, we have discussed related  collinear states that give rise to orbital active ferroaxial order.

\begin{table*}[]
\begin{tabular}{|c|c|c|}
\hline
Spin conductivity & Space Group (momentum ${\bf Q}$) & Representative Materials \\ \hline
$\sigma^l_{ij}=-\sigma^l_{ji}$ &$\begin{matrix}
123(A,M,Z), 124(M), 125(Z), 127(Z), 129(Z), 131(M),\\ 132(M), 134(A), 139(M), 140(M), 162(A), 164(A),\\ 166(T), 191(A), 221(M,X), 223(M), 225(L,X), 226(X), 227(L)  
\end{matrix}$  &$\begin{matrix}
\text{U$_2$Rh$_2$Sn,U$_2$Ni$_2$In,Ce$_2$Cu$_2$In}\\
\text{Tb$_2$Pd$_{2.05}$Sn$_{0.95}$,Yb$_2$Pd$_2$In$_{0.4}$Sn$_{0.6}$}
\end{matrix}$  \\ \hline
$\sigma^l_{ij}=\sigma^l_{ji}
$ &$\begin{matrix}
55(R,S), 56(R), 58(S), 62(U), 127(A,M), 128(M),\\ 135(M), 136(M), 138(A), 200(R), 201(R), 204(H), 206(H)
\end{matrix}$  & $\begin{matrix}
\text{ Bi$_2$Fe$_4$O$_9$,Bi$_4$Fe$_5$O$_{13}$F}\\
\text{ AgMnVO$_4$,NaFePO$_4$}\\
\end{matrix}$ \\ \hline
Mixed &$\begin{matrix}
14(D,E), 52(T), 53(R,U), 58(T,U), 60(S), 64(R),\\ 83(A,M,Z), 84(M), 85(Z), 86(A), 87(M), 128(R),\\ 136(R), 147(A), 148(T), 175(A), 202(L), 203(L)
\end{matrix}$  &$\begin{matrix}\text{Cr$_2$ReO$_6$}\\
\end{matrix}$  \\ \hline
\end{tabular}
\caption{Examples of dissipationless nonrelativistic spin conductivities and representative materials from ordering at momenta ${\bf Q}$ for various space groups. The full list containing is in the SI.}
\end{table*}

{\it Spin Conductivities:} We now study nonrelativistic spin conductivities in $J^l_i=\sigma^l_{ij}E_j$. Here, $J^l_i$ is a spin current along $i$ with spin parallel to $\hat{l}$, and $E_j$ is the electric field along $j$. Spin rotations require $\hat{l}//{\bf S}_1\times{\bf S}_2$. Real-space symmetries require $\Gamma_\text{VSC}\subset \Gamma_V\otimes\Gamma_V$, where $\{x,y,z\}\in\Gamma_V$ is the polar vector representation, describing $J^l_i$ and $E_j$.
By matching $\Gamma_V\otimes\Gamma_V$  with $\Gamma_\text{VSC}$, we can determine the allowed spin conductivities.

As a first example, consider the tetragonal space group 127 (P4/mbm) with wavevector ${\bf Q}=(0,0,\pi)$, relevant for U$_2$Rh$_2$Sn\cite{prokevs2017}, U$_2$Ni$_2$In\cite{nakotte1996}, Yb$_2$Pd$_2$In$_{0.4}$Sn$_{0.6}$\cite{martinelli2019}, Tb$_2$Pd$_{2.05}$Sn$_{0.95}$\cite{laffargue1997}, and Ce$_2$Cu$_2$In\cite{kral2025}. Here $\Gamma_\text{VSC}=A_{2g}$. Since orbital angular momentum $L_z\in A_{2g}$, ${\bf S}_1\times{\bf S}_2//\hat{z}$ is then a nonrelativistic variant of Ising SOC $L_zS_z$. In $\Gamma_V\otimes\Gamma_V$, since $\{x,y\}\otimes_A\{x,y\}=A_{2g}$, the coplanar AFM generates the nonrelativistic antisymmetric spin Hall conductivity $\sigma^z_{xy}=-\sigma^z_{yx}$, the same as that from an SOC of the form $L_zS_z$. 

As a second example, consider the orthorhombic space group 62 (Pnma) with wavevector ${\bf Q}=(\pi,0,\pi)$, relevant for AgMnVO$_4$\cite{yahia2015}, NaFePO$_4$\cite{avdeev2013}, and Ni$_2$SiO$_4$\cite{newnham1965,colman2009}. Here  $\Gamma_\text{VSC}=A_{g}$. Since none of the orbital angular momenta belong to $A_{g}$, ${\bf S}_1\times{\bf S}_2$ is different from the ${\bf L}\cdot{\bf S}$ VSC. In $\Gamma_V\otimes\Gamma_V$, since $\{x^2,y^2,z^2\}\in A_{g}$, nonrelativistic spin conductivities are purely longitudinal: $\sigma^l_{xx},\sigma^l_{yy},\sigma^l_{zz}\neq0$.  Under SOC, ${\bf S}_1\times{\bf S}_2//\hat{z}$ has the same symmetry as $L_xS_y-L_yS_x$. We therefore refer to this type of VSC as a ferroaxial VSC\cite{hayami2022ferroaxial}.

In Table I, we summarize a variety of $\Gamma_{\bf Q}$ that host purely transverse, purely longitudinal, and mixed spin conductivities, together with representative materials. The complete list of $\Gamma_{\bf Q}$ and the corresponding nonzero $\sigma_{ij}^l$ is provided in the SI.

We emphasize that the resulting spin conductivities are different from those  in 
ferromagnetic/altermagnetic systems. In the spin transport relation $J^l_i = \sigma^l_{ij} E_j$,
both the spin current $J^l_i$ and the electric field $E_j$ are T-even. The
dissipative character of $\sigma^l_{ij}$, specifically, its dependence on the relaxation time
$\tau$, is governed by the T-symmetry of the underlying order
parameter~\cite{Freimuth2014, Jungwirth2017, watanabe2024, oiwa2022}. For $T$-breaking
orders such as collinear altermagnets, $\sigma^l_{ij}$ must be an odd function of
$1/\tau$, rendering the spin conductivity dissipative. Such a spin conductivity in $T$-breaking systems is necessarily purely longitudinal as in Ohm's law (see SI for details). In contrast, the spin conductivity from $T$-preserving VSC 
can be dissipationless and can contain both transverse and longitudinal components. These spin conductivities are
purely quantum mechanical in origin, determined entirely by the band structure and coherence
factors. The expression using spin Berry curvatures is in the next section.

\noindent{\it Microscopic analysis: }

While our symmetry arguments imply the existence of dissipationless non-relativistic spin conductivities, microscopic models are needed to verify these predictions. We carry out a two-prong approach: i) initially develop general SDW models to compute spin conductivities and ii) carry out first principles calculations for the $(P,T)=(+,+)$ spin-active magnet U$_2$Ni$_2$In in which a nonrelativistic spin Hall conductivity is large.  

A general AFM Hamiltonian takes the form
\begin{equation}
\begin{split}
H&=\begin{pmatrix}
h_0({\bf k})&O_M\\O_M^\dagger&\widetilde{h_0}({\bf k+Q})
\end{pmatrix}\\
&=\begin{pmatrix}
\epsilon_0+t_0+t_x\tau_x+t_z\tau_z & {\bf J}_1\cdot\vec{\sigma}+\tau_z{\bf J}_2\cdot\vec{\sigma}\\{\bf J}_1\cdot\vec{\sigma}+\tau_z{\bf J}_2\cdot\vec{\sigma}&\epsilon_0-t_0+\widetilde{t}_x\tau_x+\widetilde{t}_z\tau_z
\end{pmatrix}
\\
\end{split}
\label{E:H}
\end{equation}
The first line follows the standard SDW framework\cite{auerbach2012}. The first block $h_0(\mathbf{k})$ is the nonmagnetic Hamiltonian, and
$\widetilde{h_0}({\bf k+Q})$ is obtained from
$h_0({\bf k})$ by shifting hopping coefficients, considering the Bloch phase factor. The second line is a general tight-binding model for space groups with Wyckoff multiplicity 2  with an on-site inversion 
center (see SI for an example with Wyckoff multiplicity 4 in space group 62 Pnma). Here, the Pauli matrix $\tau_x$ represents inter-sublattice hopping, and $\tau_z$ captures the asymmetry between the intra-sublattice hopping on the two sublattices. Time-reversal and inversion operators are $T=i\sigma_yK$ and $P=1$ (together with ${\bf k}\rightarrow -{\bf k}$). The coplanar AFM state is taken as $({\bf J}_1,{\bf J}_2)=(\frac{{\bf S}_1+{\bf S}_2}{2},\frac{{\bf S}_1-{\bf S}_2}{2})=S(\hat{x},\hat{y})$. $T$ is preserved up to one-unit-cell translation, described by the symmetry $\rho_zT$ in the AFM phase. 

\begin{figure}[h]
\centering
\includegraphics[width=7cm]{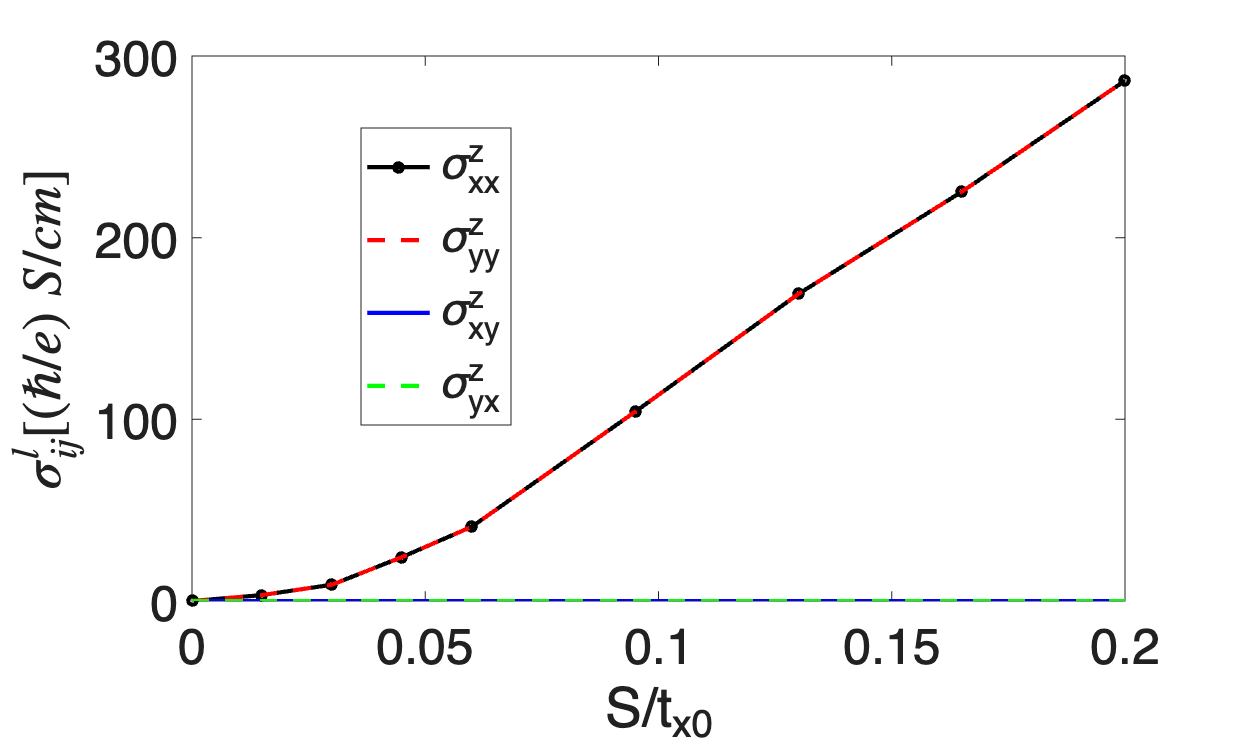}
\caption{Spin conductivities for SG127(2c) with wavevector $A=(\pi,\pi,\pi)$. The T-even order ${\bf S}_1\times{\bf S}_2$ has $A_{1g}$ spatial symmetry, giving a dissipationless longitudinal spin conductivities with $\sigma_{xx}^l=\sigma_{yy}^l$. Hopping parameters $t_{x0}=1 $eV, $t_{z0}=0.5t_{x0}$, $t_1=0.8t_{x0}$, $t_2=0$, and $\mu=0.3t_{x0}$ are used. The lattice constant is 6\AA. The nonrelativistic spin conductivity is significant for moderately strong exchange coupling.}
\label{F:3}
\end{figure}

The hopping parameters $t_{x,z}$ are governed by symmetry and depend on the specific space group and Wyckoff position, as studied for altermagnets\cite{roig2024} and included in SI. For SG 127 and Wyckoff 2c, 
the simplest hopping coefficients consistent with symmetry are $t_x=t_{x0}\cos\frac{k_x}{2}\cos\frac{k_y}{2}$, $t_z=t_{z0}\sin k_x\sin k_y$, and $\epsilon_0+t_0=t_1(\cos k_x+\cos k_y)-\mu+t_2\cos k_z$. For ${\bf Q}=(\pi,\pi,\pi)$, the shifted coefficients in $\widetilde{h_0}({\bf k+Q})$ are $\widetilde{t_x}=t_{x0}\sin\frac{k_x}{2}\sin\frac{k_y}{2}$, $\widetilde{t_z}=t_z$, and $\epsilon_0-t_0=-t_1(\cos k_x+\cos k_y)-\mu-t_2\cos k_z$. From our group theory analysis, ${\bf S}_1\times{\bf S}_2$ transforms as $A_{1g}$ under real-space symmetries, leading to a nonrelativistic, dissipationless and longitudinal spin conductivity $\sigma^l_{xx}=\sigma_{yy}^l$, as verified in Fig.\ref{F:3} using Kubo formula (See SI for details).
Figure \ref{F:3} also reveals that for a reasonable choice of parameters, it can become appreciable, for example $\sigma^z_{xx}\approx \qty{300} {(\unit{\hbar\per\elementarycharge})}\unit[per-mode=symbol]{\siemens\per\centi\meter}$. The purely longitudinal spin conductivities for space group 62 and ${\bf Q}=(\pi,0,\pi)$ can be found in SI.

\begin{figure}[htb]
\centering
\includegraphics[width=7cm]{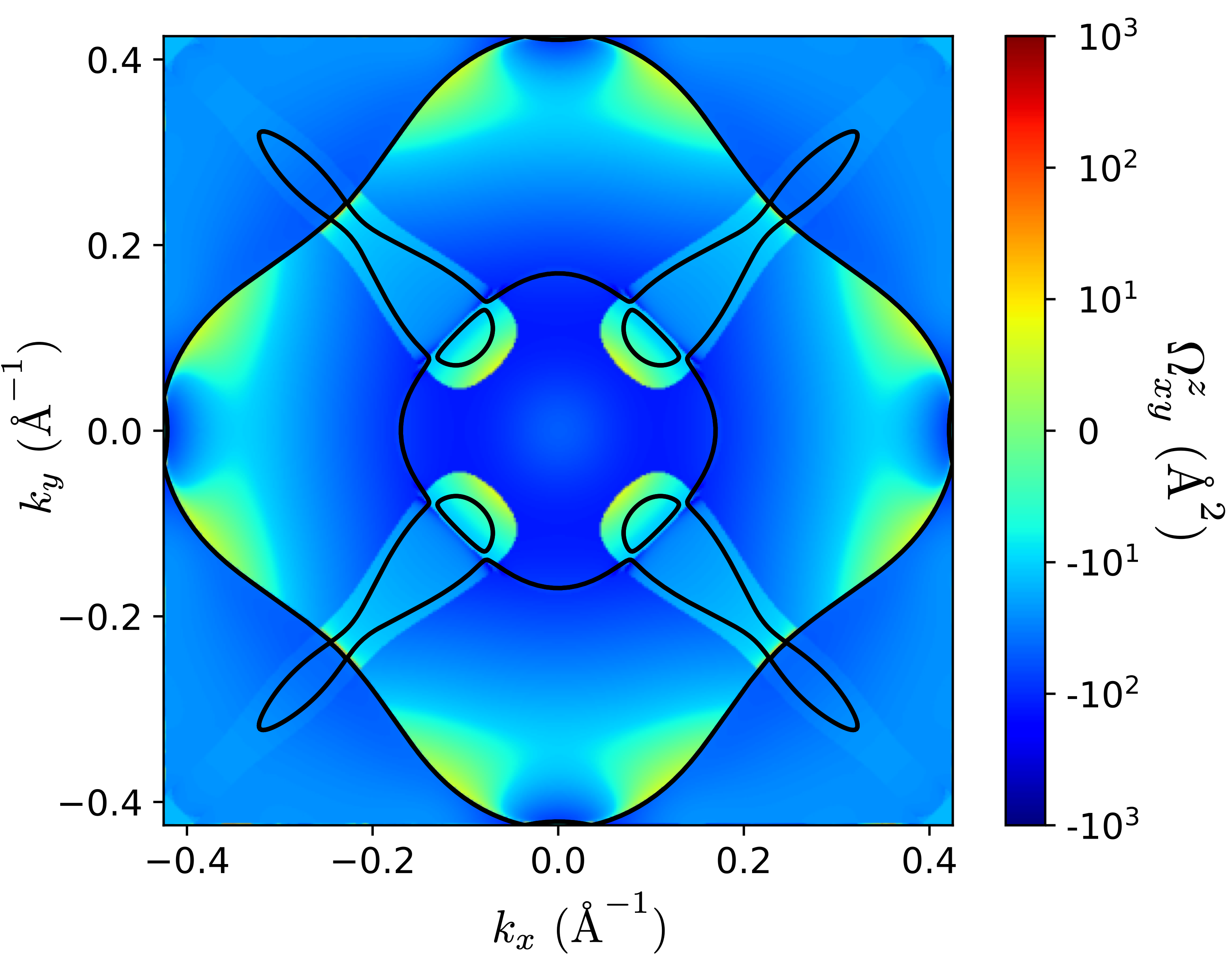}
\caption{Nonrelativistic spin Berry curvature of U$_2$Ni$_2$In summed over the occupied bands in the $k_z = 0$ plane. The solid lines indicate the intersections of the Fermi surface with this plane. The color scale is shown on a logarithmic scale.}
\label{F:4}
\end{figure}

We now turn to our DFT calculations of the nonrelativistic spin conductivity in U$_2$Ni$_2$In. Details of the calculations are described in the Supplementary Material. This compound crystallizes in the nonmagnetic space group P4/mbm (No. 127) and is known to host a coplanar AFM spin order lying within the $xy$ plane, with wave vector ${\bf Q}=(0,0,\pi)$ \cite{Martin-Martin1999-rg}. Our symmetry analysis predicts a non-zero nonrelativistic spin Hall conductivity for $\sigma^z_{xy}=-\sigma^z_{yx}$. As shown in Fig.\ref{F:4}, the DFT results for the spin Berry curvature agree with the symmetry prediction. Here, we defined the spin Berry curvature $\Omega^l_{ij}$ as 
\begin{equation}
\begin{split}
\Omega^{l}_{ij}&(\mathbf{k})
= -2\,\mathrm{Im}\sum_{m,n}
\frac{1}{(E_m-E_n)^2}\\
\times
&\left[
\langle m|\partial H/\partial k_i|n\rangle
\langle n\bigl|\{\partial H/\partial k_j, \hat{\sigma}_l /2\}|m\rangle-(i\leftrightarrow j)
\right],
\end{split}
\end{equation}
where $|m,n\rangle$ are eigenstates of $H({\bf k})$. Notably, the magnitude of the peak spin Berry curvature is comparable to that of Pt\cite{Qiao2018-co}. From these calculations, we find a substantial nonrelativistic spin Hall conductivity arising from the $(P,T)=(+,+)$ state, with $\sigma^z_{xy}=-\sigma^z_{yx}= \qty{-191.1}{(\unit{\hbar\per\elementarycharge})}\unit[per-mode=symbol]{\siemens\per\centi\meter} $. In contrast, all other nonrelativistic spin Hall components vanish, $\sigma^x_{yz}=\sigma^y_{xz}=0$, in agreement with the symmetry analysis. When SOC is included, the spin Hall conductivities become $\sigma^z_{xy}=-\sigma^z_{yx} = \qty{-417.7}{(\unit{\hbar\per\elementarycharge})}\unit[per-mode=symbol]{\siemens\per\centi\meter}$. We find that approximately \qty{46}{\percent} of the spin Hall conductivity, $\sigma^z_{xy} = -\sigma^z_{yx}$, survises even if we neglect strong SOC in uranium.

{\it Circularly polarized light: }
While $(P,T)=(+,+)$ spin-active magnets exhibit no spin splitting, here we show that such a spin-splitting can be generated by circularly polarized light (CPL).
CPL is described by a time-dependent vector potential $\vec{A}=(A_0\sin \omega t,A_0\cos \omega t,0)$. The symmetry breaking of CPL can be captured by the time-independent $T$-odd polarization $\vec{A}\times\partial_t\vec{A}$. Here, $\vec{A}$ is a polar vector $A_i\in\Gamma_V\otimes\Gamma_1^S$ ($\Gamma_1^S$ denotes a spin-scalar). The cross product transforms as $\vec{A}\times\partial_t\vec{A}\in(\Gamma_V\otimes_A\Gamma_V)\otimes\Gamma_1^S$. 
When acting on $(P,T)=(+,+)$ magnet, the CPL can induce a even-parity spin splitting $\bf M$, through the trilinear symmetry-allowed coupling $(\vec{A}\times\partial_t\vec{A}){\bf M}\cdot ({\bf S}_1\times{\bf S}_2)$. This trilinear term requires ${\bf M}$ to be T-odd, transforming as $\Gamma_M\otimes\Gamma^S_A$ with $\Gamma_M\subset(\Gamma_V\otimes_A\Gamma_V)\otimes(\Gamma_{\bf Q}\otimes_A\Gamma_{\bf Q})$, and ${\bf M}//{\bf S}_1\times {\bf S}_2$. The spin splitting is proportional to $S^2A_0^2$.

The CPL-induced spin splitting can be either ferromagnetic or altermagnetic. For example, in  tetragonal $D_{4h}$, in-plane polar vectors satisfy $(A_x,A_y)\in E_u$. The antisymmetric product $\vec{A}\times\partial_t\vec{A}$ then transforms as $(E_u\otimes_A E_u)\otimes\Gamma_1^S=A_{2g}\otimes\Gamma_1^S$. The CPL unlocks a spin-splitting for coplanar AFM at 127Z (e.g. U$_2$Ni$_2$In) as $\Gamma_M=A_{2g}\otimes A_{2g}=A_{1g}$, giving a ferromagnet. The spin-splitting for AFM at 127A (e.g. BaNd$_2$PtO$_5$) is $\Gamma_M=A_{2g}\otimes A_{1g}=A_{2g}\sim k_xk_y(k_x^2-k_y^2)$, giving a g-wave altermagnet. These symmetry arguments are supported by our microscopic Floquet theory using the minimal model introduced in Eq.~\ref{E:H} and can be found in SI.

{\it Conclusions: }
We have developed a group-theory-based framework for generating $(P,T)=(+,+)$ spin-active magnets from coplanar antiferromagnets. We constructed 166 phenomenological cases of period-doubling AFM that allow such a state and provided minimal microscopic models for some of these. Our analysis reveals the existence of non-relativistic spin conductivities in the co-planar AFM state. Our microscopic models and DFT calculations on U$_2$Ni$_2$In reveal that these spin conductivities are appreciable. We further demonstrated that circularly polarized light can induce even-parity ferromagnetic or altermagnetic spin splittings. Our theory is applicable to 16 AFM materials in Magndata database, offering a general framework for spintronics applications in realistic systems.
\bibliography{citation}

\section{acknowledgement}
D.F.A. and Y.Y. were supported by the Department of Energy, Office of Basic Energy Science, Division of Materials Sciences and Engineering under Award No DE-SC0021971.
This work is supported by Grant-in-Aid for Scientific Research from JSPS KAKENHI Grant
No.~JP26KJ1020 (J.M.),
No.~JP23K13058 (H.W.),
No.~JP24K00581 (H.W.),
No.~JP25H02115 (H.W.),
No.~JP21H04990 (R.A.),
No.~JP25H01246 (R.A.),
No.~JP25H01252 (R.A.),
JST-CREST No.~JPMJCR23O4(R.A.),
JST-ASPIRE No.~JPMJAP2317 (R.A.),
JST-Mirai No.~JPMJMI20A1 (R.A.),
and RIKEN TRIP initiative (RIKEN Quantum, Advanced General Intelligence for Science Program, Many-body Electron Systems).
H.W. was also supported by JSR Corporation via JSR-UTokyo Collaboration Hub, CURIE.
\appendix
\clearpage
\onecolumngrid
\section{Landau Theory}
Here, we present the phase diagrams for the Landau free energy in Eq.\ref{E:Landau}.
\subsection{$\beta_4=0$, $\beta_2\neq\beta_3$}
\begin{figure}[h]
    \centering
    \includegraphics[width=6cm]{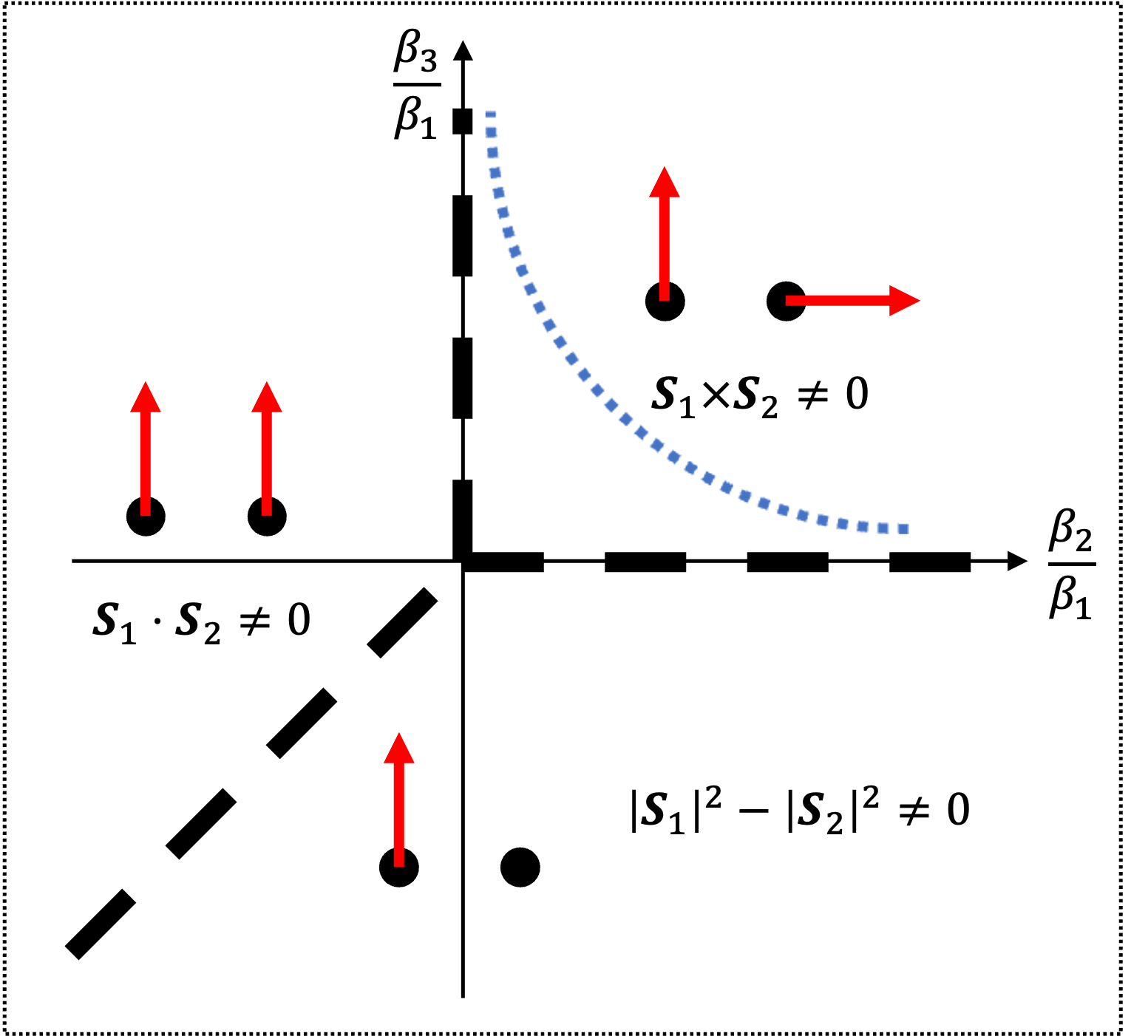}
    \caption{Phase diagram for $\beta_4=0$ is shown as thick black dashed line. When $\beta_4\neq0$, the two collinear phases mix up, and the phase boundary between collinear and coplanar phases is at the blue dotted line.  }
    \label{F:A1}
\end{figure}

The results for $\beta_4=0$ and $\beta_2\neq\beta_3$ can be found in \cite{fernandes2016,yu2025}. There are three competing phases: $({\bf S}_1,{\bf S}_2)\propto(\hat{x},\hat{x})$, $(0,\hat{x})$, and $(\hat{x},\hat{y})$. The phase diagram is shown in Fig.\ref{F:A1}. The phase boundaries are located at $\beta_2=0$, $\beta_3=0$, and $\beta_2=\beta_3$. The coplanar phase requires $\beta_{2,3}>0$.

\subsection{$\beta_4\neq0$}
When all the real-space symmetries take $({\bf S}_1,{\bf S}_2)$ to either $\pm({\bf S}_1,{\bf S}_2)$ or $\pm({\bf S}_2,-{\bf S}_1)$, inner products ${\bf S}_1\cdot{\bf S}_2$ and ${\bf S}_1\cdot{\bf S}_1-{\bf S}_2\cdot{\bf S}_2$ carry the same symmetry. This allowed $\beta_4\neq0$, and happens in 64 2D IRs, listed in the middle part of Table.II in SI.

For the phase diagram, we compare solutions with the same $R\equiv{\bf S}_1\cdot{\bf S}_1+{\bf S}_2\cdot{\bf S}_2$. The energy contributions from the quadratic term and the $\beta_1$ term are then the same. It is convenient to define $X\equiv2{\bf S}_1\cdot{\bf S}_2$, and $Y\equiv{\bf S}_1\cdot{\bf S}_1-{\bf S}_2\cdot{\bf S}_2$. $(X,Y)$ belongs to the circle $X^2+Y^2\leq R^2$. The free energy is $f=f_0(R)+\beta_2X^2+\beta_3Y^2+\beta_4XY$, where $f_0$ only depends on R.

If the energy minimum is located on the boundary of the circle, spin orderings are collinear. We can take $X=R\cos\theta$ and $Y=R\sin\theta$, and the free energy becomes $f=f_0(R)+R^2(\beta_2\cos^2\theta+\beta_3\sin^2\theta+\beta_4\cos\theta\sin\theta)$. Minimizing the energy gives $(\cos2\theta,\sin2\theta)\propto-(\beta_2-\beta_3,\beta_4)$. For generic $\beta_i$, the solution is $({\bf S}_1,{\bf S}_2)=(S_1\hat{x},S_2\hat{x})$, with $S_1\neq S_2\neq0$. The free energy is $f=f_0(R)+R^2\left[\frac{\beta_2+\beta_3}{2}-\sqrt{(\frac{\beta_2-\beta_3}{2})^2+(\frac{\beta_4}{2})^2}\right]$

If the energy minimum is located inside the circle, the solution satisfies $\partial_Xf=\partial_Yf=0$. Without fine-tuning, the only solution is $(X,Y)=(0,0)$. This corresponds to the coplanar spin ordering $({\bf S}_1,{\bf S}_2)=S(\hat{x},\hat{y})$. The free energy is $f=f_0(R)$.

The phase boundary between the collinear and coplanar phases is at $2\beta_2\beta_3=\beta_4^2$ with $\beta_{2,3}>0$ (blue dotted line  in Fig.\ref{F:A1}). As $\beta_4\rightarrow0$, the boundary recovers to $\beta_2=0$ and $\beta_3=0$.

\subsection{$\beta_2=\beta_3$}
When inner products $2{\bf S}_1\cdot{\bf S}_2$ and ${\bf S}_1\cdot{\bf S}_1-{\bf S}_2\cdot{\bf S}_2$ belong to a 2D IR $E_x$, $\beta_2=\beta_3=\beta$. $\beta_4$ is zero, since the two inner products transform differently. This happens in 40 2D IRs, listed in the last part of Table.\ref{T:1}. 

While the phase boundary between the coplanar phase and the two collinear phases remains valid $\beta_2=\beta_3=0$ (Fig.\ref{F:A1}), the quartic free energy is insufficient to determine the phase transition into the collinear states.

These 40 scenarios all carry a three-fold rotational symmetry, and $E_x\otimes_SE_x\otimes_SE_x$ contains the trivial IR. The Landau term formed by $X\equiv2{\bf S}_1\cdot{\bf S}_2$ and $Y\equiv{\bf S}_1\cdot{\bf S}_1-{\bf S}_2\cdot{\bf S}_2$ resembles those in hexagonal systems. For example, if $(X,Y)$ transform as $E_{2g}=\{2xy,x^2-y^2\}$ in hexagonal system $D_{6h}$, the free energy is:
\begin{equation}
\begin{split}
f=aR+\beta_1R^2+\beta(X^2+Y^2)+\gamma R(X^2+Y^2)
+uY(Y^2-3X^2)
\end{split}
\end{equation}

We focus on the collinear phases, where $X^2+Y^2=R^2$. The free energy simplifies to
\begin{equation}
\begin{split}
f&=f_1(R)+uY(Y^2-3X^2)
\end{split}
\end{equation}
where $f_1$ only depends on $R$. There are three solutions, describing the spontaneous three-fold rotational symmetry breaking. For $u>0$, the three solutions are $X=R\cos\theta,Y=R\sin\theta$, with $\theta=\pi/6+2n\pi/3$. For $u<0$, the three solutions are $\theta=-\pi/6+2n\pi/3$.

\section{Comparison with Ising SOC}
For collinear ferromagnets and altermagnets, as well as for Ising SOC with spin polarization
along $\hat{l}$, the spin rotational symmetry about $\hat{l}$ is preserved and $S_l$ remains
a good quantum number. In this regime, spin-$S_l = \pm$ electrons transport spin independently,
and the spin conductivity $\sigma_{ij}^l$ equals $(\hbar/2)$ times the difference in
electrical conductivities $\sigma_{ij}$ between the $S_l = \pm$ sectors. The Onsager
relations within the Kubo formalism further constrain the structure of the conductivity
tensor: electrical conductivities that are odd in $1/\tau$ must be longitudinal (as in Ohm's
law), while those that are even in $1/\tau$ must be transverse (as in the Hall effect).
Consequently, the spin conductivity arising from a ferromagnet or altermagnet satisfies
$\sigma_{ij}^l = \sigma_{ji}^l \propto 1/\tau$, whereas that from Ising SOC satisfies
$\sigma_{ij}^l = -\sigma_{ji}^l \propto \tau^0$.

The case $\mathbf{S}_1 \times \mathbf{S}_2 \parallel \hat{l}$ is qualitatively distinct: since $S_l$ is no longer a good quantum number, the spin conductivity may be either longitudinal or transverse, with the allowed components determined by the real-space symmetries. In particular, a dissipationless longitudinal spin conductivity $\sigma_{ij}^l = \sigma_{ji}^l$ is permissible in systems that preserve time-reversal symmetry. 


\section{Kubo formula}
We apply the Kubo formula to compute the dissipationless spin conductivities in Fig.3:
\begin{equation}
\begin{split}
\sigma_{ij}^l=-\frac{e}{a}\text{Im}\int_{BZ}\frac{d^3k}{(2\pi)^3}\sum_{m,n}\frac{f(E_m)-f(E_n)}{(E_m-E_n)^2}\langle m|\hat{J}_i|n\rangle\langle n|\hat{J}_j^l|m\rangle
\end{split}
\end{equation}
Current and spin current operators are $\hat{J}_i=\partial H/\partial k_i$ and $\hat{J}_j^l=\frac{1}{2}\{\partial H/\partial k_j,\sigma_l/2\}$. $f(E)=\theta(-E)$ is the zero-temperature Fermi distribution and $a$ is the lattice constant. Numerical calculations use $100^3$ grids in the Brillouin zone.

For collinear ferromagnet/altermagnet spin orderings, as well as Ising SOC along $\hat{l}$, the spin operator $S_l$ is a good quantum number. Denoting $S_l=\pm$ eigenstates as $|m^\pm\rangle$, the Kubo formula simplies to $\sigma_{ij}^l=\frac{\hbar}{2}(\sigma_{ij}^+-\sigma_{ij}^-)$, where $\sigma_{ij}^\pm$ are the electrical conductivies in the $S_l=\pm$ sectors:
\begin{equation}
\begin{split}
\sigma_{ij}^\pm\equiv-\frac{e}{a}\text{Im}\int_{BZ}\frac{d^3k}{(2\pi)^3}\sum_{m,n}\frac{f(E_{m\pm})-f(E_{n\pm})}{(E_{m\pm}-E_{n\pm})^2}\langle m^\pm|\hat{J}_i|n^\pm\rangle\langle n^\pm|\hat{J}_j^l|m^\pm\rangle
\end{split}
\end{equation}

Electrical conductivities satisfy the Onsager relations. The dissipationless components are purely transverse $\sigma_{ij}^\pm=-\sigma_{ji}^\pm$. One can check that by interchanging $(m,n)$ in the summation. To have a dissipationless longitudinal spin conductivities $\sigma^l_{ij}=\sigma_{ji}^l$, spin rotational symmetry along $\hat{l}$ thus needs to be broken.

\begin{figure}[h]
\centering
\includegraphics[width=7cm]{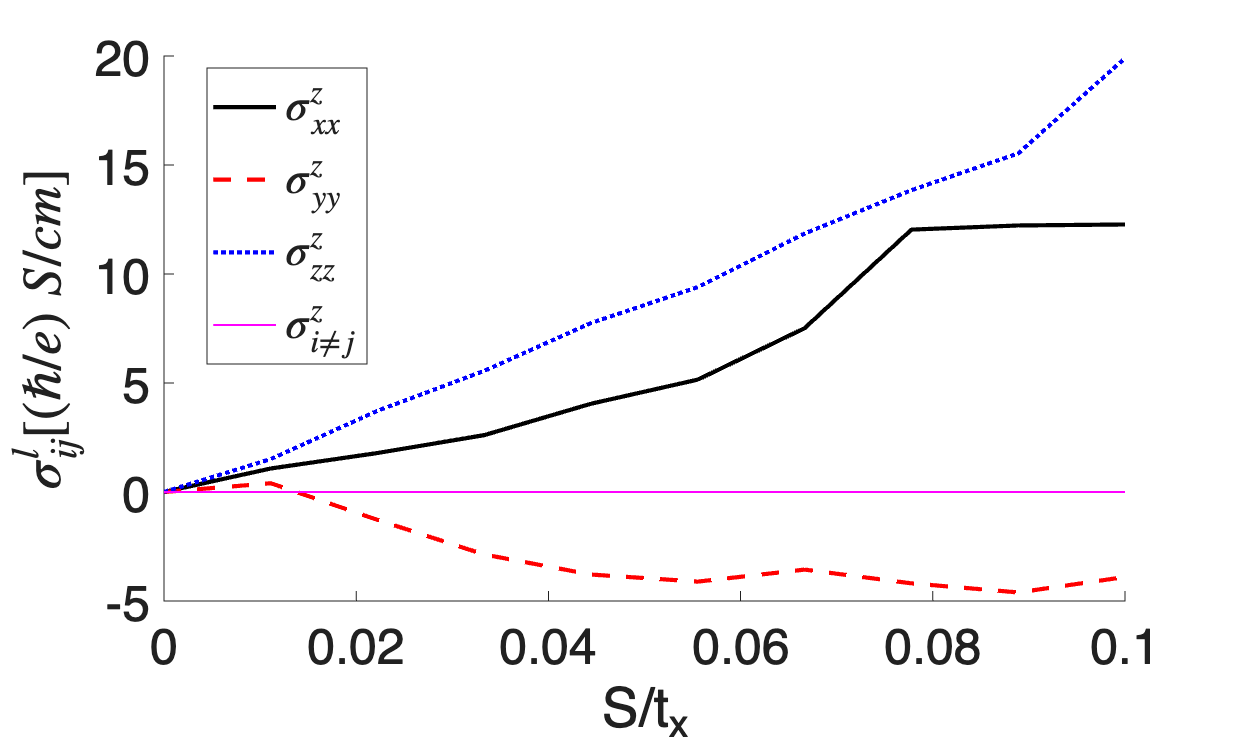}
\caption{Spin conductivities for SG62(4a) with wavevector $A=(\pi,0,\pi)$. The T-even order ${\bf S}_1\times{\bf S}_2$ has $A_{g}$ spatial symmetry, giving a dissipationless longitudinal spin conductivities with $\sigma_{xx}^l\neq\sigma_{yy}^l\neq\sigma_{zz}^l$. Hopping parameters $t_x=t_y=1$ (units in eV), $t_{1}=0.5$, $t_{2,5,6,8}=0.1$, $t_{3}=\mu=0.3$, $t_{4,9}=0.2$, and $t_7=0.05$ are used. The lattice constant is 6\AA. }
\label{F:A2}
\end{figure}

\section{Spin conductivities for space group 62 (Pnma)}
In this section, we present the modeling and calculations for space group 62 (Pnma). The nonmagnetic unit cell has a minimal Wyckoff multiplicity 4. 
For Wyckoff position (4a), the four atoms are located at $(0,0,0)$, $(0,1/2,0)$, $(1/2,0,1/2)$, and $(1/2,1/2,1/2)$ in the nonmagnetic unit cell. We use Pauli matrices $\tau_i$ and $\rho_j$ to label these sites: wavefunctions on each site have eigenvalues $(\tau_z,\rho_z)=(+,+)$, $(+,-)$, $(-,+)$, and $(-,-)$, respectively. Time-reversal and inversion operators are $T=i\sigma_yK$ and $P=1$ (together with ${\bf k}\rightarrow -{\bf k}$). Non-symmorphic spatial mirror symmetries are $\{M_x|1/2,1/2,1/2\}=\tau_x\rho_x$, $\{M_y|0,1/2,0\}=\rho_x$, and $\{M_z|1/2,0,1/2\}=\tau_x$ with the corresponding changes in momentum. 

Following the standard SDW framework in the main text (first line of Eq.2), the normal-state Hamiltonian $h_0({\bf k})$ is $8\times8$. The symmetry-allowed hopping operators in $h_0({\bf k})\equiv\sum_{ij}t_{ij,{\bf k}}\tau_i\rho_j\sigma_0$ are:
\begin{equation}
\begin{split}
&t_{00}=t_x\cos k_x+t_y\cos k_y-\mu;\; 
t_{01}=t_1\cos\tfrac{k_y}{2};\;
t_{03}=t_2\sin k_x\sin k_y;\;
t_{10}=t_3\cos\tfrac{k_x}{2}\cos\tfrac{k_z}{2};\\
&t_{11}=t_4\cos\tfrac{k_x}{2}\cos\tfrac{k_y}{2}\cos\tfrac{k_z}{2};\;
t_{13}=t_5\sin\tfrac{k_x}{2}\sin k_y\cos\tfrac{k_z}{2};\;
t_{30}=t_6\sin k_x\sin k_z;\;
t_{31}=t_7\sin k_x\cos\tfrac{k_y}{2}\sin k_z\\
&t_{33}=t_8\sin k_y\sin k_z;\;
t_{22}=t_9\cos\tfrac{k_x}{2}\sin\tfrac{k_y}{2}\sin\tfrac{k_z}{2}.
\end{split}
\end{equation}
For ${\bf Q}=(\pi,0,\pi)$, relevant for AgMnVO$_4$\cite{yahia2015}, NaFePO$_4$\cite{avdeev2013}, and Ni$_2$SiO$_4$\cite{newnham1965,colman2009}, we have $\widetilde{h_0}({\bf k+Q})=h_0({\bf k+Q})$. In the AFM phase, the spin orderings on the first two atoms are ${\bf S}_1$, and those on the last two atoms are ${\bf S}_2$ (and they can be opposite in the neighboring unit cell due to AFM). The magnetic term is $O_M={\bf J}_1\cdot\vec{\sigma}+\tau_z{\bf J}_2\cdot\vec{\sigma}$ with $({\bf J}_1,{\bf J}_2)=(\frac{{\bf S}_1+{\bf S}_2}{2},\frac{{\bf S}_1-{\bf S}_2}{2})=S(\hat{x},\hat{y})$. The spin conductivities from Kubo formula is shown Fig.\ref{F:A2}. As predicted in the group theory analysis in the main text, the spin conductivities are dissipationless and purely longitudinal with $\sigma_{xx}^z\neq\sigma_{yy}^z\neq\sigma_{zz}^z$.

\section{DFT calculations}
DFT calculations were performed using the Vienna \textit{ab initio} Simulation Package (VASP)\cite{Kresse1996-oj}. 
Projector augmented-wave (PAW) PBE pseudopotentials were employed \cite{Perdew1996-hz, Blochl1994-bi}.
The crystal and magnetic structure of U$_2$Ni$_2$In were adopted from Ref. \cite{Martin-Martin1999-rg}.
The Brillouin zone was sampled using a $\Gamma$-centered $6 \times 6 \times 8$ k-point mesh, and the plane-wave energy cutoff was set to \qty{400}{\unit{\electronvolt}}. 
The threshold of energy convergence was set to $10^{-6}$ \unit{\electronvolt}.

\begin{figure}[h]
    \centering
    \includegraphics[width=\linewidth]{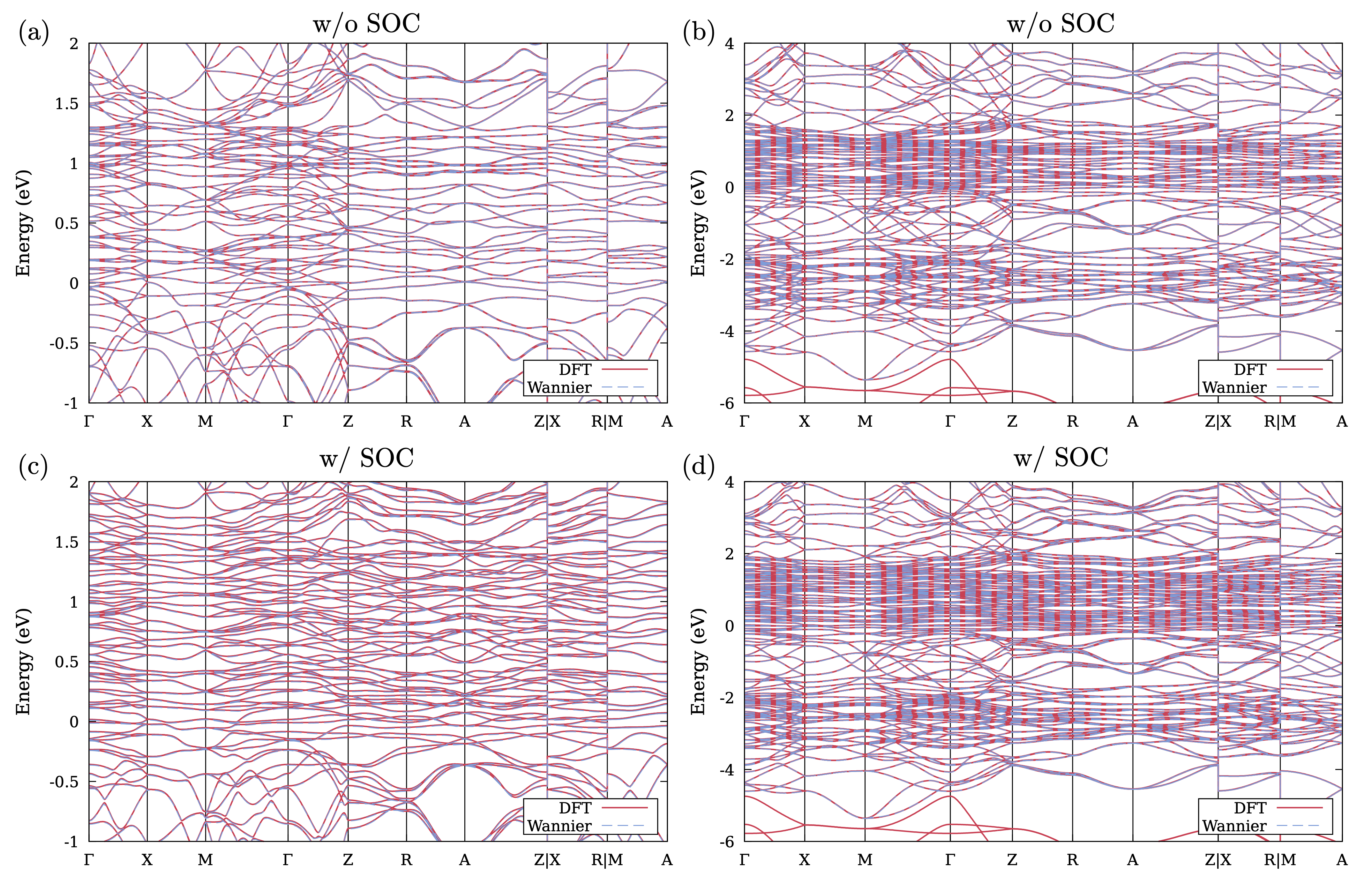}
    \caption{Band structures of U$_2$Ni$_2$In along the standard high-symmetry path of the Brillouin zone \cite{Setyawan2010-fz}. Red solid (blue dashed) lines represent DFT (Wannier-Hamiltonian) bands. The Fermi energy is set to 0. SOC was excluded in (a), (b) and included in (c), (d). The DFT bands were plotted using \textsc{VASPKIT} \cite{Wang2021-ha}.
    }
    \label{fig:supple_bands}
\end{figure}

Maximally localized Wannier functions were constructed using Wannier90 \cite{Marzari1997-gq,Souza2001-at, Pizzi2020-gd}.
For the initial projections, we included U-$d$, U-$f$, In-$p$, Ni-$s$, Ni-$p$, and Ni-$d$ orbitals. 
The resulting Wannier Hamiltonian has a dimension of $360 \times 360$, including spin degrees of freedom.
The band structures computed using the Wannier Hamiltonian are in good agreement with the DFT band structures within the energy range of -6 to \qty{3}{\unit{\electronvolt}}, as shown in Fig.\ref{fig:supple_bands}.

The spin Hall conductivity and spin Berry curvature were evaluated using \textsc{WannierBerri} \cite{Tsirkin2021-vo}.
For the Wannier interpolation, the first Brillouin zone was discretized with a $120 \times 120 \times 160$ mesh. 
The Wannier interpolation formalism for the spin Hall conductivity and spin Berry curvature follows Ref. \cite{Qiao2018-co}.
We plotted the Fermi lines utilizing \textsc{postw90.x} code.

\begin{figure}[h]
    \centering
    \includegraphics[width=\linewidth]{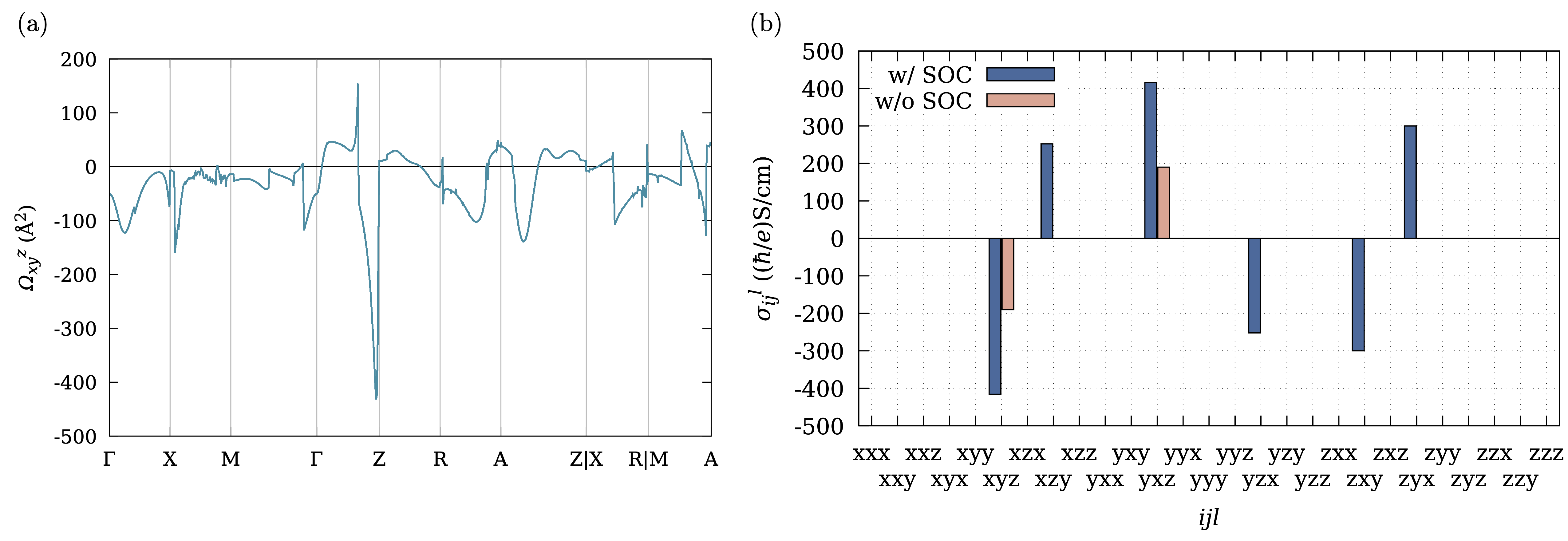}
    \caption{(b) The spin Hall conductivities with (blue) and without (red) SOC. The horizontal represents the index of spin Hall conductivities.
    }
    \label{fig:supple_compare}
\end{figure}

The spin Berry curvature summed over the occupied bands along the high-symmetry line is shown in Fig.~\ref{fig:supple_compare}(a). The peak value is approximately apporoximately \qty{-400}{\unit{\angstrom}^2}, which is comparable to those in Pt, $\alpha$-Ta and $\beta$-Ta \cite{Qiao2018-co}.
Fig \ref{fig:supple_compare} (b) compares the spin Hall conductivities obtained with and without SOC.
Without SOC, only $\sigma_{xy}^z$ and $\sigma_{yx}^z$ are nonzero, satisfying $\sigma_{xy}^z = -\sigma_{yx}^z$ in agreement with the symmetry analysis.
By contrast, when SOC is included, $\sigma_{xy}^z = -\sigma_{yx}^z, \sigma_{xz}^y = -\sigma_{yz}^x$, and $-\sigma_{zx}^y =  \sigma_{zy}^x$ become nonzero.

\section{Microscopic Floquet theory}
Microscopically, CPL leads to a time-dependent Hamiltonian, that can be analyzed using Floquet-Bloch theory\cite{rudner2020}. The light is described through Peierls substitution $H({\bf k})\rightarrow H(t)=H({\bf k+A}(t))$. The eigenstate is expressed by the Floquet-Bloch state $|\psi(t)\rangle=\exp(-i\varepsilon t)|\phi(t)\rangle$. The quasi-energy $\varepsilon $ is well-defined up to $m\omega$ for integer $m$. The Floquet state is periodic $|\phi(t)\rangle=|\phi(t+2\pi/\omega)\rangle$, and satisfies $(\varepsilon_n+i\partial_t)|\phi(t)\rangle=H(t)|\phi(t)\rangle$. In frequency space, the eigenvalue equation reads:
\begin{equation}
(\varepsilon+m\omega)|\phi^{(m)}\rangle=\sum_{m'}H^{(m-m')}|\phi^{(m')}\rangle,
\end{equation}
where $|\phi^{(m)}\rangle$ and $H^{(m)}$ is the $m$-th Fourier component of the Floquet state and the periodic Hamiltonian. 

Fourier components $h^{(\pm m)}$ contain m-th order Bessel functions, which scale as $A_0^m$ for small $A_0$. Focusing on the leading components $h^{(\pm1)}$, the effective normal-state Hamiltonian is \cite{kitagawa2011,goldman2014,mikami2016,bukov2015,eckardt2015}:
\begin{equation}
h^{(0)}_{eff}=h^{(0)}+\frac{[h^{(-1)},h^{(1)}]}{\omega}.
\end{equation}

The correction to the normal-state Hamiltonian is from the commutator between $\tau_x$ and $\tau_z$. This results in a $\tau_y$ term with coefficients 
$-2i/\omega(t_x^{(-1)}t_z^{(1)}-t_x^{(1)}t_z^{(-1)})$. Here, $t_{x,{\bf k}}^{(\pm1)}$ are the $\pm 1$ frequency components of $t_{x,{\bf k+A}(t)}$. Note that the  $\tau_y$ term is odd under time-reversal symmetry with k-even coefficients, consistent with the breaking of this symmetry with CPL.

For SG127 with Wyckoff position 2c, $t_{x,z}^{(\pm1)}$ are:
\begin{equation}
\begin{split}
&t_x^{(1)}=(t_x^{(-1)})^*=\frac{t_{x0}}{2}J_1(\frac{A_0}{\sqrt{2}})e^{-i\pi/4}(-i\sin\frac{k_x+k_y}{2}
+\sin\frac{k_x-k_y}{2})\\
&t_z^{(1)}=(t_z^{(-1)})^*=\frac{t_{z0}}{2}J_1(\sqrt{2}A_0)e^{-i\pi/4} (i\sin(k_x+k_y)+\sin(k_x-k_y))
\end{split}
\end{equation}
Here $J_1(x)$ is the first Bessel function. The effective Hamiltonian is 
\begin{equation}
h^{(0)}_{eff}=\epsilon^{(0)}_{0,{\bf k}}+t^{(0)}_{0,{\bf k}}+t^{(0)}_{x,{\bf k}}\tau_x+t_{y,{\bf k}}\tau_y+t^{(0)}_{z,{\bf k}}\tau_z,
\end{equation}
where zeroth frequency components carry zeroth-order Bessel function, e.g. $t_{x}^{(0)}=t_{x}J_0(A_0/\sqrt{2})$,  $t_{z}^{(0)}=t_{z}J_0(\sqrt{2}A_0)$. For small $A_0$, these zeroth-order Bessel functions approach 1, and we will neglect them in the following discussion. $t_{y,{\bf k}}=t_{y0}\cos \frac{k_x}{2}\cos \frac{k_y}{2}(\cos k_x-\cos k_y)\in B_{1g}$, with $t_{y0}=-2t_{x0}t_{z0}J_1(A_0/\sqrt{2})J_1(A_0\sqrt{2})/\omega$. For Wyckoff position 127(2c), operators $\tau_{y,z}$ carries the $B_{2g}$ site symmetry. The $t_y\tau_y$ term thus carries $B_{1g}\otimes B_{2g}=A_{2g}$ symmetry. It is T-odd since $\tau_y$ is imaginary. For small $A_0$, $t_{y0}$ scales as $A_0^2$. The $t_y\tau_y$ term thus follows $\vec{A}\times\partial_t\vec{A}$.

We include the $t_y\tau_y$ term in the AFM Hamiltonian to check spin splittings:
\begin{equation}
\begin{split}
&H_{eff}^{(0)}=\begin{pmatrix}
h_0({\bf k})+t_y\tau_y&O_M\\O_M^\dagger&\widetilde{h_0}({\bf k+Q})+\widetilde{t_y}\tau_y
\end{pmatrix}
\\
&\equiv  \epsilon_0+t_1\rho_0\tau_x+t_3\rho_z\tau_x+t_5\rho_0\tau_z+t_6\rho_z\tau_z+t_2\rho_0\tau_y+t_4\rho_z\tau_y+t_0\rho_z+\rho_x({\bf J}_1\cdot\vec{\sigma}+\tau_z{\bf J}_2\cdot\vec{\sigma}),
\end{split}
\end{equation}
where $t_{1,3} \equiv (t_x\pm\widetilde{t_x)}/2$, $t_{2,4} \equiv (t_y\pm\widetilde{t_y)}/2$, and $t_{5,6} \equiv (t_z\pm \widetilde{t_z})/2$. The dispersion for $t_0=0$ is:
\begin{equation}
\begin{split}
&E_{\alpha\beta\gamma}=\gamma\left\{2S^2+|{\bf t}|^2+2\beta[S^2(|{\bf t}|^2+t_5^2-t_6^2)+2\alpha S^2(t_1t_4-t_2t_3)+(t_1t_3+t_2t_4+t_5t_6)^2]^{1/2}\right\}^{1/2}
\end{split}
\end{equation}
Here, $\alpha,\beta,\gamma=\pm$ and $|{\bf t}|^2\equiv t_1^2+t_2^2+t_3^2+t_4^2+t_5^2+t_6^2$. The spin splitting depends on $S^2(t_1t_4-t_2t_3)\propto S^2(t_y\widetilde{t_x}-t_x\widetilde{t_y})$. For 127A, $\widetilde{t_y}=-t_{y0}\sin\frac{k_x}{2}\sin\frac{k_y}{2}(\cos k_x-\cos k_y)$, and the spin splitting is proportional to $S^2A_0^2\sin k_x\sin k_y (\cos k_x-\cos k_y)$. The g-wave spin splitting with magnitude $S^2A_0^2$ agrees with the group theory analysis.

\begin{table*}[htb]
\caption{$\Gamma_{\bf Q}\otimes_{A,S}\Gamma_{\bf Q}$ for 2D IRs $\Gamma_{\bf Q}$ with trivial inversion. The point group keeping TRIM ${\bf Q}$ invariant is included. Antisymmetric direct product $\Gamma_{\bf Q}\otimes_A\Gamma_{\bf Q}$ is put inside the bracket, describing the symmetry breaking from the coplanar state ${\bf S}_1\times{\bf S}_2$.  The first component, e.g. $A_g$ in $C_{2h}$, describes the trivial symmetry breaking of ${\bf S}_1\cdot{\bf S}_1+{\bf S}_2\cdot{\bf S}_2$. The other components outside the bracket denote the symmetry breaking from the collinear states ${\bf S}_1\cdot{\bf S}_1-{\bf S}_2\cdot{\bf S}_2$ and ${\bf S}_1\cdot{\bf S}_2$. When number 2 appears, such as in the $2B_g$ IRs in $C_{2h}$, the two collinear states have the same symmetry and belong to the same 1D IR (then $\beta_4\neq0$ in Landau theory). When $E_x$ IR appears, such as the $E_g$ IR in $S_6$, the two collinear states belong to the same 2D IR (then $\beta_2=\beta_3$ in Landau theory). }
\begin{tabular}{c|c}
\toprule
$\begin{matrix}
  D_{2h}:A_g+B_{ig}+B_{jg}+[B_{kg}]\\
  i\neq j\neq k
\end{matrix}$&
$\begin{matrix}
 52(T^{}_{1+,1-}),53(U^{}_{1+,1-},R^{}_{1+,1-}),\\58(U^{}_{1+,1-},T^{}_{1+,1-}),60(S^{}_{1+,1-}),128(R^{}_{1+,1-}),136(R^{}_{1+,1-})   
\end{matrix}$\\\hline

$D_{4h}:A_{1g}+B_{1g}+B_{2g}+[A_{2g}]$&
$\begin{matrix}123(Z^{}_{5+,5-},M^{}_{5+,5-},A^{}_{5+,5-}),124(M^{}_{5+,5-}),125(Z^{}_{5+,5-}),127(Z^{}_{5+,5-}),\\129(Z^{}_{5+,5-}),131(M^{}_{5+,5-}),132(M^{}_{5+,5-}),134(A^{}_{5+,5-}),139(M^{}_{5+,5-}),\\140(M^{}_{5+,5-}),221(M^{}_{5+,5-},X^{*}_{5+,5-}),223(M^{}_{5+,5-}),225(X^{*}_{5+,5-}),226(X^{*}_{5+,5-})\end{matrix}$\\\hline

$D_{4h}:A_{1g}+A_{2g}+B_{2g}+[B_{1g}]$&
$127(M^{}_{5+,5-},A^{}_{5+,5-}),128(M^{}_{5+,5-}),135(M^{}_{5+,5-}),136(M^{}_{5+,5-}),138(A^{}_{5+,5-})$\\\hline\hline\hline 

$C_{2h}:A_g+2B_g+[A_g]$&
$14(D^{*}_{1+2+,1-2-},E^{*}_{1+2+,1-2-}),64(R^{}_{1+2+,1-2-})$\\\hline

$D_{2h}:A_g+2B_{ig}+[A_g]$&
$\begin{matrix}55(S^{}_{1+2+,1-2-,3+4+,3-4-},R^{}_{1+2+,1-2-,3+4+,3-4-}),56(R^{}_{1+2+,1-2-,3+4+,3-4-}),\\58(S^{}_{1+2+,1-2-,3+4+,3-4-}),62(U^{}_{1+4+,1-4-,2+3+,2-3-})\end{matrix}$\\\hline

$C_{4h}:A_g+2B_g+[A_g]$&
$\begin{matrix}83(Z^{}_{3+4+,3-4-},M^{}_{3+4+,3-4-},A^{}_{3+4+,3-4-}),\\84(M^{}_{3+4+,3-4-}),85(Z^{}_{3+4+,3-4-}),86(A^{}_{3+4+,3-4-}),87(M^{}_{3+4+,3-4-})\end{matrix}$\\\hline

$D_{4h}:A_{1g}+2B_{2g}+[A_{1g}]$&
$\begin{matrix}127(M^{}_{1+4+,1-4-,2+3+,2-3-},A^{}_{1+4+,1-4-,2+3+,2-3-}),128(M^{}_{1+4+,1-4-,2+3+,2-3-}),\\135(M^{}_{1+4+,1-4-,2+3+,2-3-}),136(M^{}_{1+4+,1-4-,2+3+,2-3-}),138(A^{}_{1+4+,1-4-,2+3+,2-3-})\end{matrix}$\\\hline\hline\hline 

$S_6:A_g+E_g+[A_g]$&
$147(A^{}_{2+3+,2-3-}),148(T^{}_{2+3+,2-3-}),202(L^{*}_{2+3+,2-3-}),203(L^{*}_{2+3+,2-3-})$\\\hline

$D_{3d}:A_{1g}+E_g+[A_{2g}]$&
$\begin{matrix}162(A^{}_{3+,3-}),164(A^{}_{3+,3-}),166(T^{}_{3+,3-}),225(L^{*}_{3+,3-}),227(L^{*}_{3+,3-})\end{matrix}$\\\hline

$C_{6h}:A_g+E_{2g}+[A_g]$&
$175(A^{}_{3+5+,3-5-,4+6+,4-6-})$\\\hline

$D_{6h}:A_{1g}+E_{2g}+[A_{2g}]$&
$191(A^{}_{5+,5-,6+,6-})$\\\hline

$T_h:A_g+E_g+[A_g]$&
$\begin{matrix}200(R^{}_{2+3+,2-3-}),201(R^{}_{2+3+,2-3-}),204(H^{}_{2+3+,2-3-}),206(H^{}_{2+3+,2-3-})\end{matrix}$\\\hline

$O_h:A_{1g}+E_g+[A_{2g}]$&
$\begin{matrix}221(R^{}_{3+,3-}),224(R^{}_{3+,3-}),229(H^{}_{3+,3-})\end{matrix}$\\\hline
\end{tabular}
\label{T:1}
\end{table*}

\section{List of tables}
Table.\ref{T:1} contains real-space symmetry breaking $\Gamma_{\bf Q}\otimes_A\Gamma_{\bf Q}$ for ${\bf S}_1\times{\bf S}_2$, and $\Gamma_{\bf Q}\otimes_S\Gamma_{\bf Q}$ for ${\bf S}_1\cdot{\bf S}_2$ \& ${\bf S}_1\cdot{\bf S}_1-{\bf S}_2\cdot{\bf S}_2$ for 2D irreducible representations.

Table.\ref{T:2} contains real-space crystal symmetry breaking of the coplanar state ${\bf S}_1\times{\bf S}_2$ of 2D irreducible representaion $\Gamma_{\bf Q}$, obtained from $\Gamma_{\bf Q}\otimes_A\Gamma_{\bf Q}$. The resulting spin conductivities and material realizations in Magndata database are included.

Table.\ref{T:3} contains tight-binding coefficients $t_{x,z}$ for different primitive space groups with two inversion-invariant sublattices.

\begin{table*}[htb]
\caption{Real-space crystal symmetry breaking of the coplanar state ${\bf S}_1\times{\bf S}_2//\hat{l}$ of 2D irreducible representation $\Gamma_{\bf Q}$, and the resulting spin conductivities. Point groups that preserve the wavevector $\bf Q$ is included. Labels $^*$ denote symmetries not along the conventional axis. For instance, 202L hosts 3-fold rotation along (111) rather than (001) direction. Metallic materials are marked in red. In hexagonal systems $D_{3d},D_{6h},C_{6h},S_6$, the angle between x,y axes is 120$^\circ$. }
\begin{tabular}{c|c|c|c|c}
\toprule
&$\Gamma_{\bf Q}\otimes_A\Gamma_{\bf Q}$&Space group \& Wavevector&spin conductivity&materials\\ \hline
 
$D_{2h}$&$B_{1g}$&$53(R^{}_{1+,1-},U^{}_{1+,1-})$&$\sigma_{xy}^l,\sigma_{yx}^l$\\\hline

$D_{2h}$&$B_{2g}$&$52(T^{}_{1+,1-}),58(T^{}_{1+,1-}),128(R^{}_{1+,1-}),136(R^{}_{1+,1-})$&$\sigma_{xz}^l,\sigma_{zx}^l$&\text{\blue Cr$_2$ReO$_6$}\\\hline

$D_{2h}$&$B_{3g}$&$58(U^{}_{1+,1-}),60(S^{}_{1+,1-})$&$\sigma_{yz}^l,\sigma_{zy}^l$\\\hline

$D_{3d}$&$A_{2g}$&
$\begin{matrix}
162(A^{}_{3+,3-}),164(A^{}_{3+,3-}),\\166(T^{}_{3+,3-}),225(L^{*}_{3+,3-}),227(L^{*}_{3+,3-})    
\end{matrix}$&$\sigma_{xy}^l=-\sigma_{yx}^l$&$\begin{matrix}
\text{AgFe$_3$(SO$_4$)$_2$(OD)$_6$}\\
\text{KFe$_3$(OH)$_6$(SO$_4$)$_2$}\\
\text{NaFe$_3$(SO$_4$)$_2$(OH)$_6$}\\
\text{Gd$_2$Ti$_2$O$_7$}
\end{matrix}$\\\hline

$D_{4h}$&$A_{2g}$&$\begin{matrix}123(A^{}_{5+,5-},M^{}_{5+,5-},Z^{}_{5+,5-}),124(M^{}_{5+,5-}),125(Z^{}_{5+,5-}),\\127(Z^{}_{5+,5-}),129(Z^{}_{5+,5-}),131(M^{}_{5+,5-}),\\132(M^{}_{5+,5-}),134(A^{}_{5+,5-}),139(M^{}_{5+,5-}),140(M^{}_{5+,5-}),\\221(M^{}_{5+,5-},X^{*}_{5+,5-}),223(M^{}_{5+,5-}),225(X^{*}_{5+,5-}),226(X^{*}_{5+,5-})\end{matrix}$&$\sigma_{xy}^l=-\sigma_{yx}^l$&$\begin{matrix}
\text{\red U$_2$Rh$_2$Sn,U$_2$Ni$_2$In}\\
\text{Tb$_2$Pd$_{2.05}$Sn$_{0.95}$}\\
\text{\red Yb$_2$Pd$_2$In$_{0.4}$Sn$_{0.6}$}\\
\text{Ce$_2$Cu$_2$In}
\end{matrix}$\\\hline

$D_{4h}$&$B_{1g}$&$\begin{matrix}127(A^{}_{5+,5-},M^{}_{5+,5-}),128(M^{}_{5+,5-}),\\135(M^{}_{5+,5-}),136(M^{}_{5+,5-}),138(A^{}_{5+,5-})\end{matrix}$&$\sigma_{xx}^l=-\sigma_{yy}^l$&\\\hline

$D_{6h}$&$A_{2g}$&$191(A^{}_{5+,5-,6+,6-})$&$\sigma_{xy}^l=-\sigma_{yx}^l$\\\hline

$O_{h}$&$A_{2g}$&$221(R^{}_{3+,3-}),224(R^{}_{3+,3-}),229(H^{}_{3+,3-})$&$\begin{matrix}0\end{matrix}$\\\hline

$C_{2h}$&$A_g$&$14(D^{*}_{1+2+,1-2-},E^{*}_{1+2+,1-2-}),64(R^{}_{1+2+,1-2-})$&$\begin{matrix}\sigma_{xy}^l,\sigma_{yx}^l,\\\sigma_{xx}^l,\sigma_{yy}^l,\sigma_{zz}^l\end{matrix}$&$\begin{matrix}\text{Ba$_2$DyFeO$_5$, Ba$_2$YFeO$_5$}\\ \text{Ca$_2$MnWO$_6$}\end{matrix}$  \\\hline

$C_{4h}$&$A_g$&$\begin{matrix}
83(A^{}_{3+4+,3-4-},M^{}_{3+4+,3-4-},Z^{}_{3+4+,3-4-}),84(M^{}_{3+4+,3-4-}),\\85(Z^{}_{3+4+,3-4-}),86(A^{}_{3+4+,3-4-}),87(M^{}_{3+4+,3-4-})
\end{matrix}$&$\begin{matrix}\sigma_{xy}^l=-\sigma_{yx}^l,\\\sigma_{xx}^l=\sigma_{yy}^l;\sigma_{zz}^l\end{matrix}$\\\hline

$C_{6h}$&$A_g$&$175(A^{}_{3+5+,3-5-,4+6+,4-6-})$&$\begin{matrix}\sigma_{xy}^l=-\sigma_{yx}^l,\\\sigma_{xx}^l=\sigma_{yy}^l;\sigma^l_{zz}\end{matrix}$\\\hline

$D_{2h}$&$A_g$&$\begin{matrix}
55(R^{}_{1+2+,1-2-,3+4+,3-4-},S^{}_{1+2+,1-2-,3+4+,3-4-}),\\56(R^{}_{1+2+,1-2-,3+4+,3-4-}),\\58(S^{}_{1+2+,1-2-,3+4+,3-4-}),62(U^{}_{1+4+,1-4-,2+3+,2-3-})
\end{matrix}$&$\sigma_{xx}^l,\sigma_{yy}^l,\sigma_{zz}^l$&$\begin{matrix}
\text{{\blue Bi$_2$Fe$_4$O$_9$},AgMnVO$_4$}\\\text{NaFePO$_4$,Ni$_2$SiO$_4$ }
\end{matrix}$\\\hline

$D_{4h}$&$A_{1g}$&$\begin{matrix}
127(A^{}_{1+4+,1-4-,2+3+,2-3-},M^{}_{1+4+,1-4-,2+3+,2-3-}),\\128(M^{}_{1+4+,1-4-,2+3+,2-3-}),135(M^{}_{1+4+,1-4-,2+3+,2-3-}),\\136(M^{}_{1+4+,1-4-,2+3+,2-3-}),138(A^{}_{1+4+,1-4-,2+3+,2-3-})
\end{matrix}$&$\sigma_{xx}^l=\sigma_{yy}^l;\sigma_{zz}^l$&$\begin{matrix}
\text{BaNd$_2$PtO$_5$},\text{Bi$_4$Fe$_5$O$_{13}$F}
\end{matrix}$\\\hline

$S_6$&$A_g$&$\begin{matrix}
147(A^{}_{2+3+,2-3-}),148(T^{}_{2+3+,2-3-}),\\202(L^{*}_{2+3+,2-3-}),203(L^{*}_{2+3+,2-3-})    
\end{matrix}$&$\begin{matrix}\sigma_{xy}^l=-\sigma_{yx}^l,\\\sigma_{xx}^l=\sigma_{yy}^l;\sigma^l_{zz}\end{matrix}$&\text{Cu$_6$(SiO$_3$)$_6$(H$_2$O)$_6$}\\\hline

$T_h$&$A_g$&$\begin{matrix}
200(R^{}_{2+3+,2-3-}),201(R^{}_{2+3+,2-3-}),\\204(H^{}_{2+3+,2-3-}),206(H^{}_{2+3+,2-3-})    
\end{matrix}$&$\sigma_{xx}^l=\sigma_{yy}^l=\sigma_{zz}^l$&
\\\hline
\bottomrule
\end{tabular}
\label{T:2}
\end{table*}

\begin{table*}[h]
\caption{Tight-binding coefficients for primitive space groups with two atoms per unit cell at the inversion center. Abbreviation $c_i\equiv\cos k_i$, $s_i\equiv\sin k_i$, $c_{i/2}\equiv\cos \frac{k_i}{2}$, and $s_{i/2}\equiv\sin \frac{k_i}{2}$ applies. Prefactors are omitted from this table. For instance, $s_y(s_x,s_z)$ in SG14 means $t_{z1}s_ys_x+t_{z2}s_ys_z$. } 
\begin{tabular}{c|c|c}
\hline SG&$t_x({\bf k})$&$t_z({\bf k})$\\ \hline
14(2a-2d)&$c_{y/2}(s_xs_{z/2},c_{z/2})$&$s_y(s_x,s_z)$\\ \hline

53(2a-2d)&$c_{x/2}c_{z/2}$&$s_ys_z$\\ \hline

55(2a-2d)&$c_{x/2}c_{y/2}$&$s_xs_y$\\ \hline

58(2a-2d)&$c_{x/2}c_{y/2}c_{z/2}$&$s_xs_y$\\ \hline

83(2e,2f)&$\begin{matrix}c_{x/2}c_{y/2},\\s_{x/2}s_{y/2}(c_x-c_y)\end{matrix}$&$(c_x-c_y),s_xs_y$\\ \hline

84(2a,2b)&$c_{z/2}$&$(c_x-c_y),s_xs_y$\\ \hline

84(2c,2d)&$c_{x/2}c_{y/2}c_{z/2},s_{x/2}s_{y/2}c_{z/2}
$&$(c_x-c_y),s_xs_y$\\ \hline

123(2e,2f)&$c_{x/2}c_{y/2}$&$(c_x-c_y)$\\ \hline

\end{tabular}
\begin{tabular}{c|c|c}
\hline SG&$t_x({\bf k})$&$t_z({\bf k})$\\ \hline
124(2b,2d)&$c_{z/2}$&$s_xs_y(c_x-c_y)$\\ \hline

127(2a,2b)&$c_{x/2}c_{y/2}$&$s_xs_y(c_{x}-c_{y})$\\ \hline

127(2c,2d)&$c_{x/2}c_{y/2}$&$s_xs_y$\\ \hline

128(2a,2b)&$c_{x/2}c_{y/2}c_{z/2}$&$s_xs_y(c_{x}-c_{y})$ \\ \hline

131(2a,2b)&$c_{z/2}$&$(c_x-c_y)$\\ \hline

131(2c,2d)&$c_{x/2}c_{y/2}c_{z/2}$&$(c_x-c_y)$\\ \hline

132(2a,2c)&$c_{z/2}$&$s_xs_y$\\ \hline

136(2a,2b)&$c_{x/2}c_{y/2}c_{z/2}$&$s_xs_y$ \\ \hline

223(2a)&$c_{x/2}c_{y/2}c_{z/2}$&$(c_x-c_y)(c_y-c_z)(c_z-c_x)$\\ \hline
\end{tabular}
\label{T:3}
\end{table*}
\end{document}